\documentclass{elsart}
\usepackage{graphics}
\usepackage{graphicx}
\usepackage{epsfig}
\usepackage{amssymb}
\usepackage{paralist}
\begin{document}

\begin{frontmatter}

% Title, authors and addresses

% use the thanksref command within \title, \author or \address for footnotes;
% use the corauthref command within \author for corresponding author footnotes;
% use the ead command for the email address,
% and the form \ead[url] for the home page:
% \title{Title\thanksref{label1}}
\title{The Application Hosting Environment: Lightweight Middleware for
  Grid-Based Computational Science}
% \thanks[label1]{}
% \author{Name\corauthref{cor1}\thanksref{label2}}
\author{P. V. Coveney, R. S. Saksena and S. J. Zasada}
% \ead{email address}
% \ead[url]{home page}
% \thanks[label2]{}
% \corauth[cor1]{}
% \address{Address\thanksref{label3}}
\address{Centre for Computational Science, Department of Chemistry,
University College London, 20 Gordon Street, London, WC1H 0AJ}
% \thanks[label3]{}

\author{M. McKeown and S. Pickles}
\address{Manchester Computing, Kilburn Building, The University of
  Manchester, Oxford Road, Manchester, M13 9PL}

%\title{}

% use optional labels to link authors explicitly to addresses:
% \author[label1,label2]{}
% \address[label1]{}
% \address[label2]{}

%\author{}

%\address{}

\begin{abstract}
% start of abstract
Grid computing is distributed computing performed transparently across
multiple administrative domains. Grid middleware, which is meant to
enable access to grid resources, is currently widely seen as being too
heavyweight and, in consequence, unwieldy for general scientific
use. Its heavyweight nature, especially on the client-side, has
severely restricted the uptake of grid technology by computational
scientists. In this paper, we describe the Application Hosting
Environment (AHE) which we have developed to address some of these
problems. The AHE is a lightweight, easily deployable environment
designed to allow the scientist to quickly and easily run legacy
applications on distributed grid resources. It provides a higher level
abstraction of a grid than is offered by existing grid middleware
schemes such as the Globus Toolkit. As a result the computational
scientist does not need to know the details of any particular
underlying grid middleware and is isolated from any changes to it on
the distributed resources. The functionality provided by the AHE is
`application-centric': applications are exposed as web services with a
well-defined standards-compliant interface. This allows the
computational scientist to start and manage application instances on a
grid in a transparent manner, thus greatly simplifying the user
experience.  We describe how a range of computational science codes
have been hosted within the AHE and how the design of the AHE allows
us to implement complex workflows for deployment on grid
infrastructure.
\end{abstract}
\end{frontmatter}

% main text
\section{Introduction}
We define grid computing as distributed computing performed
transparently across multiple administrative domains
\cite{pvc2005_1,boghosian2005}.  By grid computing, we refer to any
activity involving digital information - computational processing,
visualisation, data collection from instruments and computational
analyses, database access, storage and retrieval \cite{pvc2005_1} -
performed by utilising computational, visualisation, network and
storage resources available on a grid. Grid computing has immense
potential for computational scientists.  A grid composed of resources
belonging to different institutions has the potential to provide a
level of service and support that cannot be expected from
intra-institutional resources or even an existing set of independent
resources which are not integrated into a grid as such. Some of the
benefits possible from using grid infrastructure include capacity for
increased workload volumes, faster response times/turn-around times
for the user, and increasing the frequency of running
analyses. Integration of resources into a grid infrastructure allows
for better consolidation and workload management and may result in
reduction of overall costs for grid resource owners. Grids can be used
to solve computational science problems that could not be solved by a
single resource either at all or efficiently, possibly due to compute,
memory, data or network limitations
\cite{chin2006_1,boghosian2006_1}. A grid provides a flexible and
dynamic infrastructure based on open standards which scales
efficiently as the number of component resources increases.  A general
computational grid should provide easy access to computational,
visualisation, storage, network and other resources, enabling
computational scientists to pick and choose those required to achieve
their widely varying scientific objectives.
\newline
\newline
Many emerging grids are in operation around the world, for example,
the UK National Grid Service (NGS) \cite{ngs}, US TeraGrid
\cite{teragrid}, Enabling Grids for E-sciencE (EGEE) \cite{egee} and
Distributed European Infrastructure For Supercomputing Applications
(DEISA) \cite{deisa} in the EU and the Japanese National Research Grid
Initiative (NAREGI) \cite{naregi}, which make capability for
computational processing, data storage and visualisation on high speed
networks available to computational scientists. The engagement of
scientists in grid activities is essential for such grids to mature
into genuine production-level infrastructure which can deliver the
much heralded potential of grid computing to the international
computational science community.

\section{Motivation for the Application Hosting Environment}
Grid architecture consists of various components including grid
middleware, application programming interfaces (APIs), software
development kits, protocols, services, system software and hardware
\cite{foster2001}. In our efforts over the past four years to utilise
grids to address new challenges in computational science and
engineering, we have experienced significant barriers, particularly
when dealing with heavyweight grid middleware
\cite{pvc2004_1}. Transparency, implying minimal complexity for
computational scientists using grid technology, has been missing from
existing distributed computing infrastructures that might aspire to
call themselves grids. In this paper, we present the approach we have
taken to address and overcome the heavyweight grid middleware problem.
\newline
\newline
Grid middleware is the software layer that transforms distributed,
heterogeneous resources spanning multiple administrative domains into
a single, integrated grid so that the heterogeneity and multiplicity
of the distributed resources is transparent to the user. The user
interacts with all the heterogeneous resources in the grid through a
uniform interface. Thus, middleware is a critical component of grid
infrastructure and its lack of usability has been a major barrier to
the uptake of grid technology. Many of the current grid middleware
solutions can be characterised as what we describe as `heavyweight',
that is they display some or all of the following features:
\begin{inparaenum}[(i)]
\item the grid middleware's client-side software, which is what the
users/computational scientists have to interact with on their
desktops, is difficult to install and configure,
\item has non-negligible dependencies on supporting software \item
requires non-standard ports to be opened within client-side firewalls
and
\item the server-side components of the grid middleware exhibit the
same heavyweight characteristics as the client-side software; in most
cases the situation is worse on the server-side\end{inparaenum}. As a
result, the user faces significant obstacles in deploying and using
many of the current grid middleware solutions. This has led to
reluctance amongst many scientists to actively embrace grid technology
\cite{pvc2004_1}.  It is important for the development of grid
computing as a whole that as many diverse groups of scientists as
possible begin to use grids; it is crucial that the uptake of grid
technologies occurs beyond specialized grid-centric projects
\cite{blake2005,jha2005_1,boghosian2006}. While some progress has been
made in the field of grid middleware technology \cite{globus,unicore},
the prospect of a heterogeneous, on-demand computational grid as
ubiquitous as the electrical power grid is still a long way off.
\newline
\newline
To address these deficiencies, there is now much attention focused on
`lightweight' middleware solutions
\cite{kewley2005,pvc2005_2,rixon2005,blower2005}, which attempt to
lower the barrier of entry for grid users. Our efforts to address
these deficiencies have been concentrated on the development of the
Application Hosting Environment (AHE), a lightweight WSRF \cite{wsrf}
compliant, web services based environment for hosting scientific
applications. The AHE alleviates many of the problems posed by
heavyweight grid middleware for computational scientists, who are
among those who stand to benefit the most from grid computing.  The
AHE allows scientists to quickly and easily run unmodified application
codes on grid resources, managing the transfer of files to and from
the grid resources and allowing the user to monitor the status of
application instances that are run on the grid. In the AHE, we adopt a
standards-based web services approach to expose scientific
applications on the grid as stateful web services.
\newline
\newline
From a computational scientist's perspective, a web service is simply
any computational application functionality that is accessible over a
network as a service, that can be invoked in multiple contexts,
possibly to form higher-level services. The description of the
services provided by the AHE is standards-compliant \cite{wsdl} so
that they can be invoked using any web services compliant client (see
Appendix \ref{s:arch} for more details). The AHE design is influenced
by our previous work on WEDS (WSRF-based Environment for Distributed
Simulation) \cite{pvc2005_2}, a hosting environment designed for
operation primarily within a single administrative domain. The AHE
differs from WEDS in that it is designed to operate seamlessly across
multiple administrative domains, in a true grid sense.

\section{Design of the Application Hosting Environment}
The AHE is an ensemble of programs written in Perl with Java-based
command-line and Graphical User Interface (GUI) client tools. The
purpose of the AHE is to provide a mechanism for deploying
applications onto computational grids that makes it easy for the
scientist to start and manage applications on the grid.  The AHE has a
client-server architecture in which as much of the complexity as
possible is moved to the AHE server. This makes the AHE client
thinner, with a reduced set of software dependencies, and easy to
install. Our approach works better than the conventional and
relatively inflexible portal approach as the AHE client is not
constrained to being a web browser. The AHE client has been
implemented as a desktop GUI application as well as an interoperating
set of command-line tools, making it highly flexible and powerful as a
tool for constructing lightweight workflows via scripting as described
in Section \ref{s:aheworkflows}. Furthermore, all state persistence
occurs on the AHE server, which substantially increases client
mobility.
\newline
\newline
To ensure that the AHE client is easy to install, configure and use,
whilst providing maximum functionality, a number of design constraints
were set on the overall design of the AHE:
\begin{enumerate}
\item We do not require the user to install Globus \cite{globus},
Unicore \cite{unicore} (or any other grid middleware) clients on
his/her machine even if the grid, that he wishes to run applications
on, uses such grid middleware.
\item We assume that the client device uses NAT (Network Address
Translation \cite{nat}) and that the device is firewalled to only
allow outgoing connections. This effectively means that the client
does not accept any inbound connections: all communication is
out-bound and initiated by the client.
\item For maximum portability, we require that the client only
 supports HTTP \cite{http}, HTTPS \cite{https} and SOAP \cite{soap}.
\item The client does not have to be installed on a single machine;
the user can move between clients on different machines and access the
applications that have been launched from any one of them. The user
can even employ a combination of different clients - for example, a
command line client to launch an application and a GUI client to
monitor it. The client therefore must maintain no information about an
application instance's state. All state information is maintained as a
central service on the AHE server that is queried by the AHE client.
\item The client machine needs to be able to upload input files to and
download output files from a grid resource, but we assume it does not
have GridFTP client software installed. A WebDAV-based intermediate
file staging area is therefore used to perform file-staging between
the client and the target grid resource.
\item The AHE client maintains no knowledge of the location of the
application on the target grid resource, or how it should be run, and
it maintains no information on specific environment variables that
need to be set.
\item The client should not be affected by changes to a remote grid
resource, for example if its underlying middleware changes to a newer
version of the Globus Toolkit \cite{globus,gt4}.
\item All communication is secured using Transport Layer Security
(TLS) \cite{tls} with the user's X.509 certificate \cite{x509} used
for encryption, authentication and authorization.
\end{enumerate}
These design constraints have led us to an architecture in which the
AHE client is extremely lightweight and simple to deploy, with no
dependencies except for the Java Runtime Environment \cite{jre}. It
should be noted that this design does not remove the need for
middleware solutions such as Globus or Unicore on the grid resource
(which may be, for example, a US TeraGrid site or a supercomputer on
the DEISA grid); indeed, we provide an interface to run applications
on multiple resources with different underlying grid middlewares, so
it is essential that the grid resource provides a supported middleware
installation on their machines. The AHE uses GridSAM \cite{gridsam}, a
job submission and monitoring software, as its interface to the grid
resources. GridSAM provides the AHE server with a uniform interface to
the different type of grid middleware installed on a grid. See Section
\ref{s:arch} and Figure \ref{F:arch-fig} for a brief discussion of
GridSAM's role in the AHE architecture. The AHE removes the
requirement to install any other middleware on the user's client
machine: the user simply needs to install the lightweight AHE client
to interact with the grid resources which may have heavyweight
middleware installed on them. The AHE client currently provides a
uniform interface to grid resources with installations of Globus
Toolkit 2.4.3 \cite{globus}, Grid 
Engine 6.0 u4+ \cite{sge}, Condor
v6.4, 6.6 and 6.7 \cite{condor} and Unicore \cite{unicore} while
work is in progress to enable interfacing to Globus Toolkit 4.0 \cite{gt4}
based grids. The recently developed Unicore support for AHE 
fulfils demand for the capability to launch applications on the EU DEISA
grid \cite{deisa} and brings
with it the much sought after interoperability between Unicore-based grids
and Globus-based grids.
\newline
\newline
In addition to the state information about all the application
instances, the AHE server also stores all the necessary information
about how an application should be run on the various computational
grid resources and provides a single, uniform interface to the AHE
client for running that application across all possible grid
resources. This is particularly useful for applications that run on
supercomputers which often have unique deployment scenarios and
require special runtime environments. Storing this
deployment/configuration information on the AHE server as a central
application-specific service is an efficient mechanism for making the
information available and useful to a large number of users at once.
\newline
\newline
The AHE design is based around the notion that very often a group of
researchers will want to run the same application, but not all of them
will possess the skill or inclination to install the application on a
remote set of grid resources. In the AHE, we therefore distinguish
between experts and end-users. An expert user installs the application
and configures the AHE server, so that all participating users can
share the same application on the grid at large.  The end-user simply
needs to download and install the lightweight AHE client and is then
able to trivially access such `centrally installed' applications.
\newline
\newline
The design of the AHE is novel in its application-centric approach,
according to which we treat the computational `application' as a
higher level entity than a computational `job'. Computational
steering applications \cite{fowler2005,pickles2005}, coupled model
simulations \cite{defabritiis2006,pvc2006_3} and workflows
\cite{pvc2006_1,pvc2006_2,fowler2006_2} are cases where applications
and jobs can be clearly distinguished. In the initial AHE release,
applications and jobs stand mainly in a one-to-one relationship except
for the case of workflow applications. Complex workflow applications
have been deployed on grids using the AHE wherein the workflows are
composed of simpler computational jobs. This is discussed in more
detail in Section \ref{s:aheworkflows}. By providing a service that
will launch a particular application rather than a generic
computational job it is possible to reduce the complexity of the
client and make the scientist's life easier. The scientist can then
concentrate on science rather than spending time understanding and
installing middleware and managing individual jobs.

\section{Application Hosting}
\label{s:hostingapps}
An application is said to be grid-enabled when it is able to run on
multiple heterogeneous resources comprising a grid. In the
computational science domain, the AHE provides a mechanism to expose
an unmodified application as a standards-compliant web
service. Appendix \ref{s:webservices} and \ref{s:arch} provide more
technical detail about how this is done.
\newline
\newline
When an application is launched on a grid resource using the AHE, that
instance of the application is referred to as the application
instance. In other words, the user can launch many application
instances for an application which is hosted in the AHE.  Any kind of
modelling and simulation application can be hosted within the AHE.  To
the best of our knowledge, all of the applications currently hosted in
the AHE are simulation codes as discussed in Section
\ref{s:hostingapps}. We therefore use the term simulation instead of
application instance in much of the following discussion of the AHE.
However, the same discussion is applicable, more generally, to
instances of any other type of application hosted in the AHE.
\newline
\newline
We currently host a number of scientific applications within the AHE
including the highly scalable molecular dynamics (MD) codes DL\_POLY
3.01 \cite{smith1996}, NAMD \cite{kale1999}, LAMMPS
\cite{plimpton1995} and GROMACS \cite{gromacs} and the
lattice-Boltzmann (LB) code, LB3D
\cite{chin2006_1,harting2005,lb3d}. Scientists can use the AHE client
to launch these applications on a grid; in particular, we currently
use the AHE to run these applications on the UK NGS and US TeraGrid
resources.  The AHE does not require the application to be modified as
long as the application is reasonably well-behaved with respect to the
run parameter specification, file-naming and file-location
conventions.
\newline
\newline
We next describe, how to run such applications on a grid
using the AHE. Most of the steps involved are common to all
applications. Then we discuss some of the specific applications
currently hosted in the AHE. Lastly, we describe how the AHE client
and server can be configured to host a new application.

\subsection{Running hosted applications}
\label{s:runningapps}
The steps involved in running an application instance on a grid using
the AHE are listed below. We present the steps in terms of running
application instances of simulation type applications. The AHE GUI
client is implemented in the familiar `wizard' fashion; each step in
the launching process is presented to the user as a separate screen in
the GUI with controls to navigate between screens:
\begin{enumerate}
\item In the AHE client wizard, the user first specifies the
particular application that he wishes to run, e.g., DL\_POLY, and
other constraints such as the number of processors on which to run it,
maximum wall time required and so on. The AHE server returns a list of
grid resources on which the application is installed and that match
the constraints. The user then selects the grid resource on which he
wants to run the simulation.
\item Next, the AHE client wizard prompts the user to specify the
location of the input files and the names of the output files that
would be produced at the end of the simulation. In the most general
case, the user can manually specify the locations of input files and
the names of output files. However, the AHE client has a plug-in
parser feature, whereby application-specific plug-in parsers can be
integrated into the client automating the file management
operations. This is discussed in more detail in Sections
\ref{s:exampleapps} and \ref{s:newapps}
\item The user then uses the AHE client wizard to start the simulation
on the grid resource.
\item Once the simulation has started, the user can manually check the
simulation status or choose either to set the AHE client to poll the
status of the simulation or to shut-down the client and return at a
later time to retrieve the simulation status. The user can use any
machine with an AHE client installation, not necessarily the one from
which the simulation was started to monitor his/her simulations.  This
is ideal in situations where the user would like to access the
simulation state, including the input and output files, from different
machines.
\item Apart from monitoring the status of the simulation, users can
also terminate their simulation before normal completion using the AHE
client.
\item Finally, when the simulation has completed, the user can
transfer all the input and output files on the remote grid resource at
the click of a button. The task of recovering output files scattered
around a (global) grid has been very tedious until now and this
feature of the AHE has greatly enhanced the productivity of grid
users.
\item The user may then destroy all memory of the simulation on the
AHE server or can allow the simulation state to persist on the AHE
server to review it in the future. Note that a review may involve
retrieving the simulation input and output files at a later time.
\end{enumerate}

The combination of the capability to parse configuration files in
order to discover input and output files, to automatically stage the
files to and from the grid resources, and to review the state of the
simulation including associated files long after the simulation has
finished, makes the AHE an extremely powerful tool for addressing the
challenge of solving the provenance problem especially when one wishes
to run a large number of application instances distributed across a
grid.

\subsection{Example applications}
\label{s:exampleapps}
\textbf{DL\_POLY}\newline
DL\_POLY \cite{smith1996} is a parallel
molecular dynamics package for simulations of macromolecules,
polymers, ionic systems and solutions. We host DL\_POLY within the AHE
for users who wish to run molecular dynamics (MD) simulations with
DL\_POLY on the UK NGS core nodes and the HPCx facility \cite{hpcx}.
This makes it trivial to launch DL\_POLY simulations on these grid
resources using the AHE client wizard.
\newline
\newline
A DL\_POLY specific plug-in parser has been integrated into the AHE
client, so that the user simply needs to specify the location of the
CONFIG, file on his/her local machine and all the input files, such as
CONTROL, FIELD and/or TABLE, etc., automatically get staged to the
remote grid resource at the start of the simulation.
\newline
\newline
Once the simulation has terminated, the user can, at the click of a
button, retrieve all the input and output files from his/her DL\_POLY
simulations, for example, the OUTPUT, REVCON, REVIVE, STATIS files to
the local machine.

\textbf{NAMD and LAMMPS} \newline
We host NAMD and LAMMPS within the
AHE for users who wish to run MD simulation with these applications on
the UK NGS core nodes and US TeraGrid sites.  NAMD \cite{kale1999} is
a parallel molecular dynamics code primarily used for large-scale
bio-molecular simulations. LAMMPS \cite{plimpton1995} is a parallel
molecular dynamics code with optimizations for long-range
interactions.  It is trivial to launch NAMD and LAMMPS simulations
using the AHE client wizard by following the steps described
previously.
\newline
\newline
Plug-in parsers for NAMD and LAMMPS have already been integrated into
the AHE client and are supplied with the AHE client download. These
plugin-parsers allow the user to specify the location of the
NAMD/LAMMPS configuration file from which the AHE automatically
discovers and transfers all the input and output files that the
application instance will consume and produce.
\newline
\newline
For example, the NAMD/LAMMPS configuration file along with the
force-field paramete file, co-ordinate file, velocity file, and
restart files that may be required to run the simulation are
automatically staged to the grid resource at the start of the
simulation.  Since the simulation is run within a single working
directory on the remote grid resource, any relative paths in the
configuration files are removed. All input and output files that
belong to a particular simulation are located in a uniquely associated
working directory. Once the simulation has finished the user can
again, at the click of a button, retrieve all the output files from
his/her simulation working directory on the remote grid resource.

\textbf{GROMACS}\newline
GROMACS \cite{gromacs} is a highly optimized,
parallel molecular dynamics simulation package extensively used for
biomolecular simulations. We host GROMACS within the AHE for users who
wish to run MD simulations with GROMACS on the UK NGS core nodes. A
client parser plug-in makes it trivial to launch the application using
either the AHE GUI or command-line clients.

GROMACS differs from most of the other applications currently hosted
in the AHE in that it consists of two separate applications, namely a
preprocessor (grompp) and the actual simulation code (mdrun). A Perl
script has been created which runs on the grid resource and directs
the output of the preprocessor to the input of the simulation
code. This script is then hosted in the AHE and treated as a single
application. A client configuration parser has also been created which
takes its input from a `meta' configuration file, allowing the user to
specify parameters for both components of the application.

Once complete, the AHE client stages back both the output from the
simulation code and certain important intermediate files created by
the preprocessor.

\textbf{LB3D}\newline
LB3D \cite{lb3d} is a massively parallel
implementation of the lattice-Boltzmann model for amphiphilic fluid
dynamics \cite{chin2006_1,harting2005} which is able to reproduce the
morphological and rheological phenomena observed in ternary
amphiphilic mixtures from purely bottom-up mesoscopic interactions.
We have hosted LB3D within the AHE to run lattice-Boltzmann
simulations on the UK NGS and US TeraGrid nodes.  In this case, it was
necessary for us to modify the application, as some features of the
code were incompatible with the AHE design. LB3D produces output files
whose names contain a randomly generated string while relying also on
a pre-exisiting output directory at the start of the simulation. In
order to be easily hosted within the AHE, it is preferable to know in
advance the names of the output files that the code will generate
during the execution so as to automate the process of output file
retrieval. Moreover, in the AHE design, each independent execution of
an application is associated with a unique working directory on the
remote grid resource, within which the simulation is run and the
staging and de-staging of input and output files occurs. It is,
therefore, desirable for the application code to run within a single
working directory. In grid environments, where because of the
multiplicity of resources, one needs to keep track of output dispersed
across various remote grid resources, it is particularly useful to
have working directories associated with specific simulations.

\subsection{Hosting a new application}
\label{s:newapps}
To host an application within the AHE, the expert-user needs to
configure the AHE server with the following application-specific
information:
\begin{enumerate}
\item location of the application executable on the grid resources,
e.g., NGS nodes and TeraGrid sites.
\item any environment variables that may be required to run the
application on the different grid resources such as the path to
dynamically linked libraries
\end{enumerate}
The reader should consult the AHE server installation guide
\cite{aheserverguide} for the exact location where this information
needs to be stored. After configuring the AHE server with the above
information, running a short script provided with the server
distribution updates the registry of hosted applications to include
the new application.  There is no need to restart the AHE server once
a new application has been added in this manner.
\newline
\newline
Once the above changes have been made to the server-side, the user is
able to run the application using the AHE client in the generic mode,
i.e. the AHE client does not require any modification when a new
application is hosted. However, it is possible to write an
application-specific plug-in parser for the AHE client which can parse the user-specified
application configuration file to automatically discover the input and
output files for the particular application instance that the user
wishes to launch. The parsing capability allows the user to simply
specify the location of the configuration file; the AHE client then
automatically parses it to find out all the input file locations and
output file names. Thus, when the simulation is started the input
files are automatically staged to the remote grid resource and the
output files are retrieved at the end of the simulation. Details of
writing and integrating such a plug-in parser with the AHE client
are specified in the AHE client user guide \cite{aheclientguide}.

\section{Scientific Workflows}
\label{s:aheworkflows}
In addition to the GUI, the AHE client includes a closed set of atomic
command-line tools that replicate the essential operations in the AHE
GUI client. For example, there are AHE command-line tools to `prepare'
an application instance for running on a grid resource, to `start',
`monitor', `terminate' and `getoutput(files)'. Complex workflows
composed of multiple simulations and/or calculations can be realised
very simply by writing scripts that combine the command-line tool
functionality. Here, we present three examples of how the AHE
command-line tools can be combined to compose and deploy complex
workflows.

\subsection{Ensembles of simulations}
\label{s:ensembles}
\begin{figure} [!h]
	\begin{center}
	    \includegraphics[scale=0.3]{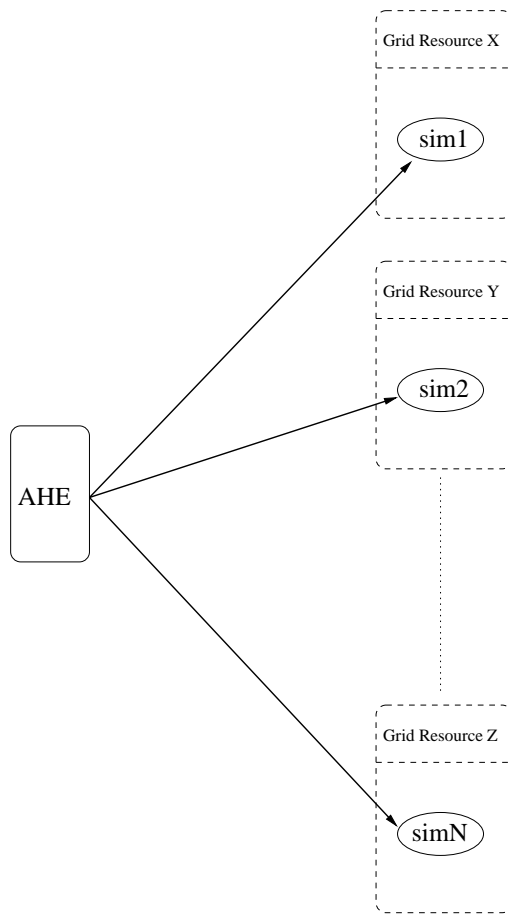}
	\end{center}
\caption{Running ensembles of application instances on a grid using
the AHE. Application instances labelled as sim1, sim2,..., simN are
started on computational resources on the grid using the AHE. The AHE
manages input/output file-staging for each application instance
to/from the associated grid resource. The user is able to monitor the
status of all the application instances.}
\label{F:ensemble-fig}
\end{figure}
In the workflow depicted in Figure \ref{F:ensemble-fig}, the user
wishes to launch a large number of independent simulations on the grid
and retrieve the results for post-processing. There may be an
additional requirement for these simulations to run
concurrently. Despite their multiplicity, each simulation may be
computationally expensive. For high-end applications, the capabilities
needed can only be delivered by resources available on grids such as
the UK NGS and US TeraGrid.  Computational resources on the grid
provide a flexible, low cost alternative to intra-institutional
supercomputing resources for running such ensembles of simulations.
Provided sufficient computational resources are available on the grid,
deployment of the ensemble workflow via the AHE results in shorter
turn-around time and greatly enhanced control of provenance for the
simulation output data as compared to submitting each simulation
individually on a local computational resource.
\newline
\newline
The ensemble functionality is currently being employed to run an
ensemble of molecular dynamics simulations of the HIV-1 protease and
inhibitors \cite{pvc2006_1}, each with a different starting
configuration for the system, in order to obtain better
statistics. Within the ensemble of simulations, the initial conditions
are identical for the simulations except for the initial atomic
velocity distributions, sampled from a Maxwell-Boltzmann distribution
and randomised differently for the same system temperature.
\newline
\newline
Such ensembles of MD simulations are useful for thermodynamic
integration (TI) calculations where multiple MD simulations need to be
launched at different values of the dual topology parameter $\lambda$
or where multiple short trajectory simulations can be launched on a
grid to give insight into the thermodynamics of a single long
trajectory simulation \cite{shirts2001}. An alternative way to study
TI calculations via chained simulations is discussed in Section
\ref{s:chains}.
\newline
\newline
Thus the AHE provides a simple way to write scripts in order to start
simulations on multiple grid resources. The user needs only to specify
the configuration file for each simulation; the AHE automatically
discovers and stages input/output files to/from the remote grid
resource without the user having to install any complicated grid
middleware on his/her local machine. This feature is particularly
useful when there are a large number of simulations to be launched on
several resources which span multiple administrative domains.
\newline
\newline
The AHE server maintains a history of all the simulations that have
been launched by each user and allows for their monitoring and review
in the future. As noted previously, the user has the flexibility to
change between client machines while still having a reference to
his/her entire simulation history.

\subsection{Chained simulations}
\label{s:chains}
\begin{figure} [!h]
        \begin{center}
           \includegraphics[scale=0.5,width=\textwidth]{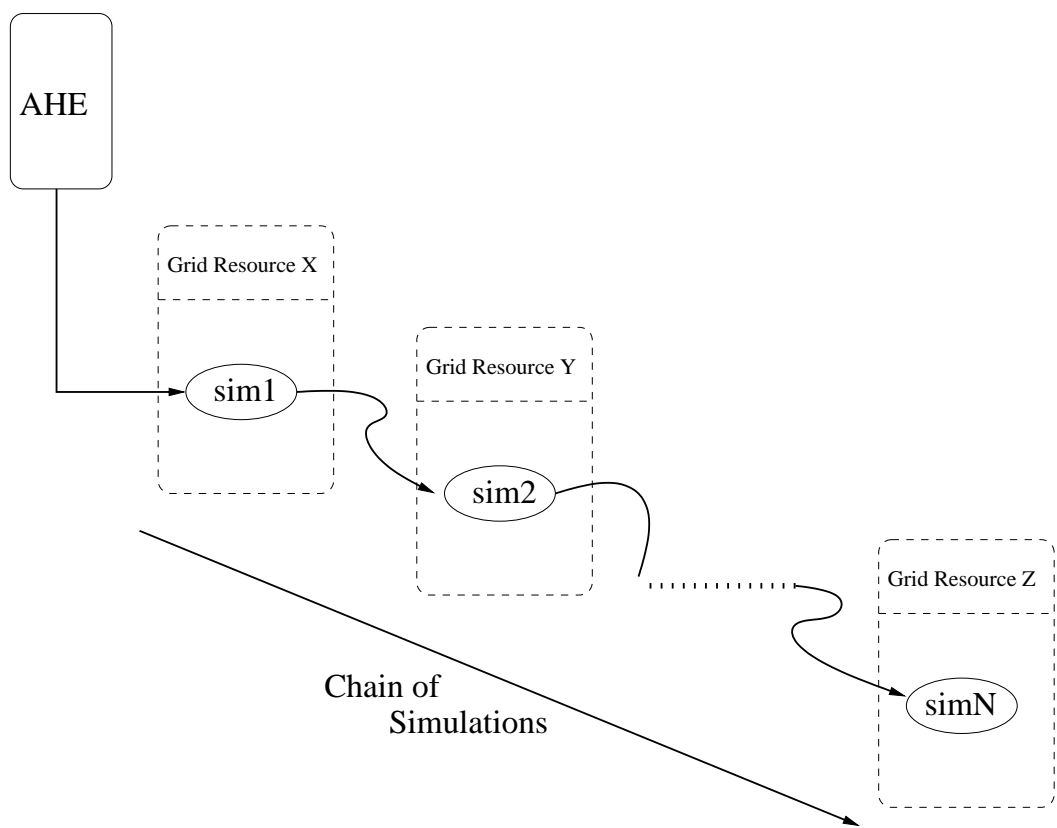}
        \end{center}
\caption{Running a chain of application instances on a grid using the
AHE.  Application instances labelled as sim1, sim2,...., simN are
started in sequence on computational resources on the grid using the
AHE. The AHE manages input/output file-staging to/from the associated
grid resources; piping the output from one application instance as
input for the next application instance in the chain. The user is able
to monitor the status of each application instance as the workflow
progresses.}
\label{F:chain-fig}
\end{figure}
\begin{figure} [!h]
        \begin{center}
           \includegraphics[scale=0.5,width=\textwidth]{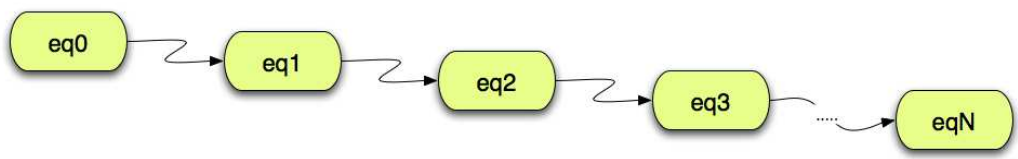}
        \end{center}
\caption{Running a chain of MD simulations on a grid; each simulation
		corresponds to a different step in the initial
		relaxation of the system to equilibrium before
		production runs can be performed.}
\label{F:chainmd-fig}
\end{figure}

In the workflow in Figure \ref{F:chain-fig}, the user wishes to launch
a sequence of chained simulations where one application instance
begins execution after the previous instance has finished and may
depend on output from the previous instance's execution.
\newline
\newline
Such a workflow of chained simulations is currently being used by
numerous AHE users. In particular, the chained simulation capability
is being used to study binding affinities of wild-type and
mutant HIV-1 proteases with drug inhibitors \cite{pvc2006_2}. In order
to obtain a starting structure for the MD production runs, the initial
artificially mutated HIV-1 protease wildtype crystal structure is
equilibrated by subjecting it to a chain of equilibration simulations,
each simulation corresponding to a step in the equilibration
procedure. This is illustrated in Figure \ref{F:chainmd-fig}.
The AHE is used to automate the entire process of launching and
managing the equilibration simulations in sequence on the grid, thus
realising the workflow in Figure \ref{F:chainmd-fig}.  The automation
provided by AHE includes starting each simulation on different grid
resources in sequence, transferring input files to the grid resource
at the start and retrieval of output files at the end of each
simulation. This feature is particularly valuable as it frees the user
from the mundane task of keeping track of all the simulation data
files and manually transferring them across the grid resources for the
next simulation in the chain.
\begin{figure} [!h]
        \begin{center}
           \includegraphics[scale=0.5,width=\textwidth]{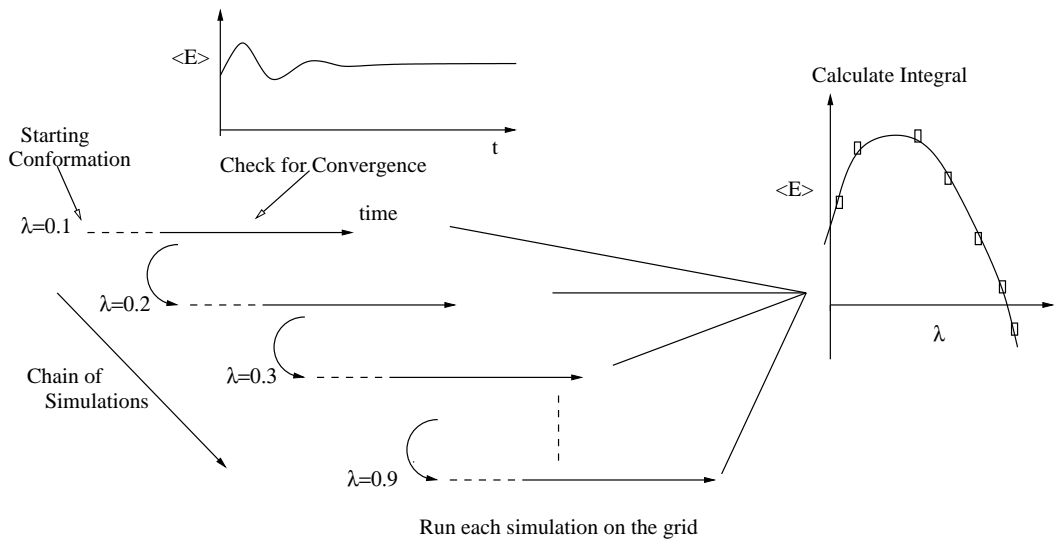}
        \end{center}
\caption{Thermodynamic integration calculation consisting of a chain
	of simulations at different values of parameter $\lambda$ is
	an ideal candidate for grid deployment.  See text in Fowler
	\emph{et al.} \cite{fowler2005} for more details.}
\label{F:ti-fig}
\end{figure}
Thermodynamic integration (TI) calculations
\cite{fowler2005,fowler2006_2}, where each simulation is run at a
particular value of the dual topology parameter $\lambda$ ($0 \leq
\lambda \leq 1$), such that the initial configuration for a simulation
at $\lambda_n+1$ is obtained from a simulation at $\lambda_n$, is
another example of a workflow of chained simulations that can be
deployed via the AHE. This is illustrated in Figure \ref{F:ti-fig}. As
noted in Section \ref{s:ensembles}, TI-type calculations can also be
studied using an ensemble workflow. Additionally, one can imagine a
chained workflow as in Figure \ref{F:chain-fig} comprising of
application instances belonging to different application types.

\subsection{Concurrent simulations}
\begin{figure} [!h]
        \begin{center}
           \includegraphics[scale=0.4]{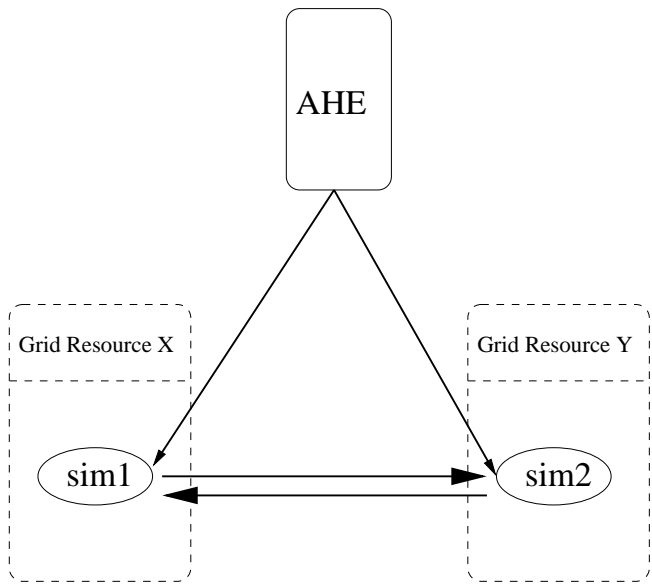}
        \end{center}
\caption{Running coupled application instances on a grid using the
AHE. Each application instance is executed on a separate grid resource
via the AHE. The AHE can monitor the individual components as they
execute.}
\label{F:coupled-fig}
\end{figure}
In the workflow in Figure \ref{F:coupled-fig}, the user wishes to
launch concurrent and dependent simulations on multiple grid
resources. The simulations may need to communicate with each other
periodically during their execution. This type of workflow can be used
to perform coupled model simulations \cite{pvc2006_3} where the system
being simulated spans multiple time-scales and/or length-scales. In
order to accurately and efficiently simulate such systems, the
different domains of the system are simulated at different levels of
detail using distinct applications which communicate at regular
intervals.  See De Fabritiis \emph{et al.} \cite{defabritiis2006} for
a description of such a multiscale computational technique that has
successfully coupled molecular dynamics with Landau's fluctuating
hydrodynamics to simulate sound waves propagating in bulk water and
reflected by a lipid monolayer.  Although such coupled models have not
been deployed with the AHE yet, the AHE provides the infrastructure to
start up the different applications on the grid, and once
communication has been established and concurrent execution has begun,
the AHE will enable the user to monitor individual applications and/or
to terminate them, while at the same time furnishing a high-level
overview of the entire coupled model application.
\newline
\newline
At the time of writing, there is a significant AHE user base with
others planning to use it. Favourable experiences have been reported
for NAMD and LAMMPS applications hosted within the AHE as compared to
alternative strategies \cite{globus,pvc2005_2,pickles2005} for running
jobs on grids \cite{pvc2006_1}. As expected, from the design criteria,
the key benefits for users have been:
\begin{compactenum}[(i)]
\item minimal installation, configuration and maintenance effort on
the client-side;
\item flexibility of restarting or switching between clients so that
the client need not be `attached' to the grid in order to launch or
manage jobs;
\item automatic staging in of input files to a grid resource, third
party file transfer between resources, and retrieval of distributed
files to the local machine on job completion;
\item no need to repeatedly submit individual jobs to many resources
and monitor status across multiple grid resources.
\end{compactenum}
Users have expressed the need for the capability to monitor queues on
different grid resources. This capability can be provided by the AHE
in future if the queue status
information is published via a web service by the resource managers
installed on the grid resources.

\section{Future Developments}
Future extensions of the AHE aim to add functionality that enables
even more ambitious computational science projects to be undertaken on
grids with ease. Globus Toolkit 4 (GT4) \cite{gt4} and Unicore \cite{unicore} 
are able to process jobs specified using the JSDL
schema and we are exploring the possibility of future versions of AHE
being able to
directly submit jobs to Globus Toolkit 4 (GT4) and Unicore
without the need for GridSAM (see Appendix \ref{s:arch}) as a
third party tool to provide 
the web service interface for job submission. Work is underway to integrate
the WSRF-compliant RealityGrid steering framework into the AHE
\cite{pickles2005}, which will provide support for starting and
managing steerable applications and coupled model applications hosted
in the AHE. We plan to extend the current workflow functionality to
orchestrate even more complex workflows, using the
industry-standard BPEL language \cite{bpel}. The scientist will 
be provided with graphical tools to interface with the BPEL 
workflow engine and easily implement their desired BPEL workflow 
operations.
\newline
\newline
We are now looking at new platforms on which the lightweight AHE
client can be deployed. For example, work is underway on an AHE client
that runs on mobile phones and PDAs \cite{kalawsky2005} allowing
scientists to launch and monitor simulations on the move. Co-allocation
is one of the major challenges faced by computational scientists
trying to perform cross-site runs using multiple concurrently
available grid resources \cite{boghosian2006}. No automated advanced
reservation and co-scheduling systems are in place yet within
so-called production grids; performing cross-site runs
\cite{boghosian2006} today requires human intervention for booking
resources in advance and making sure these are indeed available at the
desired time\footnote{Note that the beta version of the NAREGI
\cite{naregi} software stack has a super-scheduler at the heart of its
architecture \cite{naregi2} and claims to implemenent WS-Agreement
\cite{wsagreement} compliant co-allocation.}.  A fault-tolerant web
services compliant approach to co-allocation called HARC
\cite{maclaren2006} has recently been proposed and implemented. The
attractiveness of this scheduler stems from its ability to co-allocate
\begin{inparaenum}[(i)] \item compute resources and \item lambda
networks \cite{smarr2006}, which are dynamically provisioned optical
networks for bandwidth intensive applications\end{inparaenum}. An
interface to the HARC system will be developed in the AHE client to
provide support for automated, fault-tolerant co-allocation. We hope
to provide support for MPICH-G2 \cite{mpichg2} enabled applications in
the future.  Finally, we are also looking into the use of the AHE for
accessing campus-based resources, often referred to as `campus
grids'. Ultimately, AHE could become a uniform interface to all
computational resources.

\section{Conclusions}
Our goal in this work has been and continues to be to provide
scientists with tools such as command-line interfaces that are simple
and familiar, while at the same time furnishing access to a much more
powerful set of resources on grids than previously available, allowing
scientists to use their own creativity to achieve increasingly complex
computational objectives. The AHE allows computational scientists to
overcome the barrier of heavyweight, difficult-to-use grid middleware
and thereby to take advantage of grid computing with much greater ease
than has been possible hitherto. Grid computing has contributed
significantly to the scientific programme within isolated,
grid-centric computational science projects
\cite{chin2006_1,blake2005,jha2005_1,chin2004,jha2006_1} but it still
remains under-utilised among wider national and international
computational science communities. The involvement of increasing
numbers of computational scientists in grid activities is essential
for determining best practices in grid scheduling policies,
provisioning/orchestration policies and resource configuration. We
hope that the AHE will help to realise these goals.

\section{Distribution and Installation}
The first release of the AHE was made available on 31 March 2006. The
AHE (Version 1.0.1) client and server software can be downloaded from
the RealityGrid website at http://www.realitygrid.org/AHE or the
NeSCForge website at http://forge.nesc.ac.uk/projects/ahe. Future
releases will continue to be available from the RealityGrid
website. Documentation, including installation and configuration of
the AHE to run applications on the UK NGS and US TeraGrid, is also
available from these websites. Subscription to the AHE mailing list is
open at
http://www.mailinglists.ucl.ac.uk/mailman/listinfo/ahe-discuss.
\newline
\newline
A version of the AHE software, Version 1.0.2, has also been 
developed in which the AHE
server can be hosted within the Apache Jakarta Tomcat container
\cite{tomcat}. This version of the AHE is distributed via the Open
Middleware Infrastructure Institute UK (OMII-UK) \cite{omii}
within the latest OMII 3.2.0 server and client software 
which was released on 12 November 2006. This 
version of the AHE conforms to the OMII Integration
Specification and is easily installable and integrated with the rest
of the OMII distribution including WSRF::Lite \cite{wsrflite} and
GridSAM \cite{gridsam} which are pre-requisites for the AHE. OMII 3.2.0
server and client software is available for download at http://www.omii.ac.uk.

\section{Acknowledgements}
We are grateful to many current AHE users for suggestions and comments
which have improved its usability. We thank Dr. Phil Fowler, Kashif
Sadiq, Mary-Ann Thyveetil and Dr. James Suter for providing scientific
use cases during the initial development of the AHE.  In particular,
we acknowledge Kashif Sadiq for providing use cases for the ensemble
and chained simulation workflow functionality of the AHE. We would
like to thank Dr. Dorothy Duffy and Alex Rutherford for providing
expertise and use cases for hosting DL\_POLY in the AHE.  The AHE is
funded by EPSRC through the RealityGrid (GR/R67699), RealityGrid
Platform Grant (EP/C536452/1) and Rapid Prototyping of Usable Grid
Middleware (GR/T27488/01) projects, and through the OMII Managed
Programme project Robust Application Hosting using WSRF::Lite
(http://www.omii.ac.uk/mp/mp\_wsrf\_lite.jsp).
% The Appendices part is started with the command \appendix;
% appendix sections are then done as normal sections
% \appendix
% \section{}
% \label{}
\appendix
\section{Web Services and the Application Hosting Environment}
\label{s:webservices}
In the AHE, we have adopted the Service Oriented Architecture (SOA)
approach to providing applications as services on the grid. Web
services are one way of realizing SOA. As mentioned in the main text,
from a computational scientist's perspective a web service is simply
any computational functionality accessible over a network as a
service. The term `service' implies that the user simply needs to know
the description of the service in order to invoke it and does not need
to know how the functionality of the service is actually implemented;
this is particularly important on a heterogeneous grid. Hence, in
order for the web service to be consumed by a client, it is of utmost
importance that the web services have standards-compliant
interfaces. In the AHE, we use WSRF::Lite \cite{wsrflite}, a perl
library that provides the framework for writing web services.  In this
Service Oriented Architecture (SOA) approach, the key concept is that
of loosely coupled components, i.e., web services, that interact via
SOAP messages and whose interface is described by documents conforming
to the WSDL standard \cite{wsdl}.

There are obvious benefits to the adoption of a standards-based,
dynamic and flexible approach to running applications on heterogeneous
grid resources and managing the state of the application instances in
a uniform, integrated way. Such an approach allows flexible
integration of multiple applications to accomplish complex
computational tasks on the grid, as we aspire to do with the AHE's
workflow functionality. Also, the widespread adoption of the web
services approach in industry and academia permits grid computing
practitioners to take advantage of numerous existing web services and
web services development tools that are independently being made
available to the community and to take advantage of the significant
standardization and inter-operability efforts which clearly also
address significant issues in grid computing.

In the AHE, an application instance is represented as a transient
stateful web service Resource WS-Resource \cite{wsrf}. The WS-Resource
properties associated with the application instance include
\begin{itemize}
\item the application instance's reference handle or the EndPoint
Reference(EPR)
\item status of the application instance
\item a trivial name to refer to the simulation
\item date and time when the application instance was started,
\item the grid resource on which its running,
\item names and URLs of the input files
\item names and URLs of the output files.
\end{itemize}
Each time an application is run on the grid, a WS-Resource \cite{wsrf}
is created on the AHE server-side, and is used to represent that
instance of the application's execution. This WS-Resource provides an
interface for the user to interact with the application instance.  The
WS-Resource corresponding to the application instance and its
WS-Resource properties are stored on the AHE server in a database
referred to as the `App Instance Registry'.  This is described in more
detail in Appendix \ref{s:arch}. The WS-Resource properties can be
queried at any time using the AHE client or any other web services
compliant client. In this way the AHE provides a uniform interface for
managing simulations of various scientific application codes deployed
on multiple grid resources.

The WS-Resource persists even after the application instance has
finished executing, providing information on the location of any
output files as well as the input files and configuration parameters
used to initially run the application. This is a powerful provenance
capability of the AHE, as the user can return at a later time and
review the simulation by querying the properties of the associated
WS-Resource.

\section{AHE Architecture}
\label{s:arch}
\begin{figure} [!h]
        \begin{center}
           \includegraphics[scale=0.5,width=\textwidth]{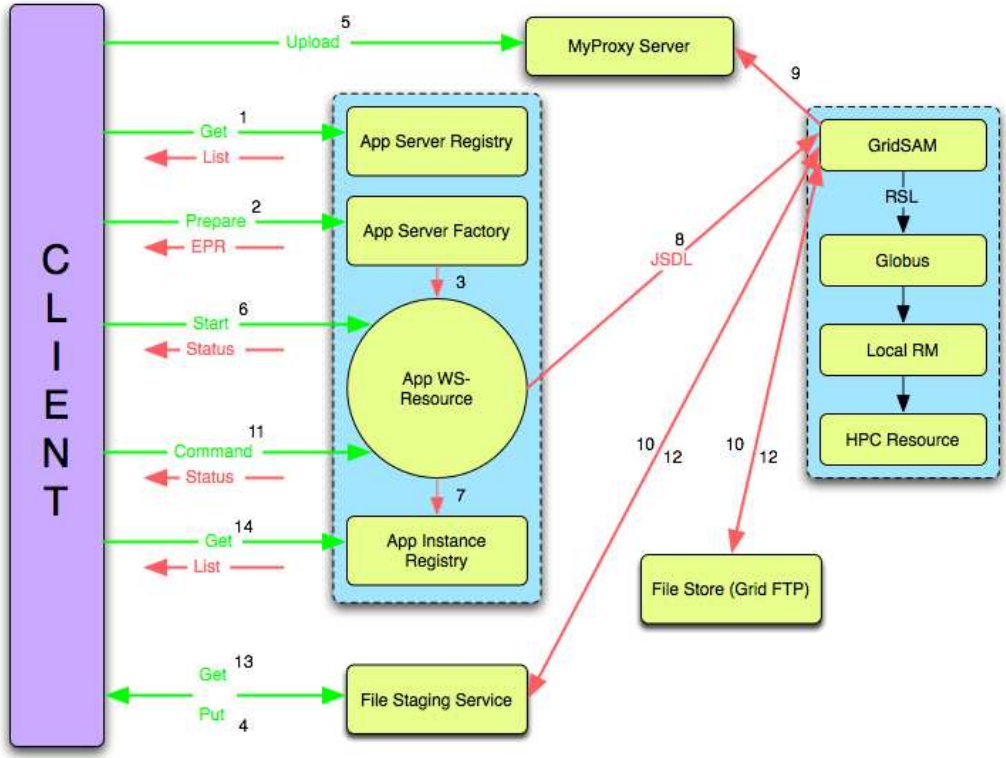}
        \end{center}
\caption{Architecture of the Application Hosting Environment; the
	 numbers denote operations performed between the AHE client,
	 AHE server, File Staging Service, File Store, MyProxy server
	 and Grid Resources. See the text in Appendix \ref{s:arch} for
	 a detailed description of this diagram.}
\label{F:arch-fig}
\end{figure}
\textbf{WSRF::Lite}
\newline The AHE is developed with WSRF::Lite, a
Perl implementation of the Web Service Resource Framework (WSRF
\cite{wsrf,wsrfglobus}) specification which has been ratified by the
OASIS \cite{oasis} standards body. WSRF::Lite is the follow-on from
OGSI::Lite, the Perl implementation of the Open Grid Service
Infrastructure (OGSI \cite{tuecke2003}) specification from GGF
\cite{ggf}.  It is built on SOAP::Lite \cite{soaplite} the Perl module
for web services from which it derives its name.

In the current release of the AHE, we host the WS-Resources on the
server-side within the Apache web server and store the WS-Resource
properties in a PostgreSQL \cite{postgresql} database.  The PostgreSQL
\cite{postgresql} database storing the simulation's state can be
replicated and backed up non-locally to provide fault tolerance.

\textbf{GridSAM} \newline
The AHE currently accesses computational
resources on the grids via middleware called GridSAM
\cite{gridsam}. GridSAM provides a web services interface, running in
an OMII \cite{omii} web services container, for submitting and
monitoring jobs on grid resources with various Distributed Resource
Managers (DRMs) - Globus 2.4.3 \cite{globus}, Grid Engine 6.0
u4+ \cite{sge} and Condor v6.4, v6.6 and v6.7 \cite{condor} - via its DRM
connector-based plugin-in architecture. GridSAM has a plug-in
architecture that allows connectors for different types of DRMs to be
integrated into it, thereby adding the functionality to submit and
monitor jobs on resources with those DRMs.

Computational jobs submitted to GridSAM are described using another
emerging standard, the Job Submission Description Language. GridSAM
consumes Job Submission Description Language (JSDL \cite{jsdl})
documents and performs the task of `job' submission to the target grid
resource.

\textbf{File Staging Area (FSA)} \newline
The `File Staging Area'
(FSA) is designed to allow the client to use the WebDAV \cite{webdav}
file transfer protocol to `GET' (step 13 in Figure \ref{F:arch-fig})
or `PUT' (step 4 in Figure \ref{F:arch-fig}) a file on the WebDAV
server so that it can be downloaded by GridSAM on to the target grid
resource (step 10/12 in Figure \ref{F:arch-fig}).  This is designed to
handle the case where input files exist on the user's client machine
and these need to be transferred to the grid resource. Similarly, the
FSA is useful when files need to be retrieved from the remote grid
resource to the local client machine. The FSA provides an intermediate
file-staging area, albeit one which is accessible via normal HTTP GET
and PUT mechanisms and is useful for pulling down the simulation
input/output files from anywhere on the grid using nothing more than a
web browser or potentially a PDA. We consider this type of file
transfer as `pass by value' since the client actually transports the
file to the FSA.

\textbf{FileStore} \newline
For much larger input/output files, which
the user may not wish to download to his/her local machine, the AHE
supports third-party file transfer using a `FileStore' mechanism. The
FileStore is any place on the grid where files required by the
instance of the application are available through GridFTP or HTTP. The
FileStore is used to hold large files like checkpoints that would not
normally be stored on the client machine. The AHE client passes the
URL \cite{url} of the source input file or target output file to the
AHE server and the input files automatically get staged from the
FileStore to the grid resource on which the simulation is to be run or
the output files from the grid resource automatically get staged out
to the desired target FileStore. These are steps 10 and 12 in Figure
\ref{F:arch-fig}. We call this `pass by reference' Note that FileStore
may be a grid resource providing a storage service such as the data
nodes on the UK NGS.

\textbf{MyProxy} \newline
Security is critical in any grid computing
environment \cite{pvc2005_1}.  We chose not to use WS-Security
\cite{mls,mls2} for the AHE because message level security is not
required: we do not have any intermediaries for relaying SOAP
messages. In the AHE, communication is via HTTPS to provide Transport
Layer Security (TLS \cite{tls}); mutual authentication is used between
the AHE client and the AHE server using X.509 digital certificates
\cite{x509}. For users of the UK NGS \cite{ngs}, digital certificates
or e-Science certificates can be obtained from the national
certification authority \cite{ukca}. Within the AHE, the App
WS-Resource is given access to a proxy certificate \cite{x509} stored
on a MyProxy \cite{myproxy} server. The AHE server implements
fine-grained authorization in that only the owner of the simulation
has access to its properties.  This can be extended to provide group
access or open access analogous to the UNIX file system permissions
for collaborative work.

\textbf{App Server Registry} \newline
The `App Server Registry'
maintains a registry of all the applications that are hosted within
the AHE, such as, DL\_POLY \cite{smith1996}, NAMD \cite{kale1999},
LAMMPS \cite{plimpton1995}, GROMACS \cite{gromacs} and LB3D
\cite{chin2006_1,harting2005,lb3d}. The user can query the registry
(step 1 in Figure \ref{F:arch-fig}) to find the address for the
service factory that provides a particular application.

\textbf{App Server Factory} \newline
The `App Server Factory' is a web
service based on the Factory Pattern \cite{factorypattern} which
creates a new App WS-Resource on the AHE server each time the user
invokes the Prepare operation.

\textbf{App WS-Resource} \newline
The `App WS-Resource' exists on the
AHE server and is a WS-Resource representing a particular application
instance. The user invokes the Prepare operation of the App Server
Factory whenever an application instance needs to be launched on the
grid. For each invocation of the Prepare operation, a new App
WS-Resource is created on the AHE server and its WS-Resource
properties are initialized as per the application instance parameters
specified by the user.

\textbf{App Instance Registry} \newline
The AHE server maintains an
`App Instance Registry' which contains the history of all application
instances/WS-Resources that the user has launched/created. The App
Instance Registry is implemented by a PostgreSQL \cite{postgresql}
database. The user can query the App Instance Registry to get a list
of all the application instances launched and the associated
WS-Resource properties. Collaborative analysis of simulation histories
is greatly aided by the stored WS-Resource properties that can be
accessed online via the App Instance Registry.

We now briefly describe the numbered operations in Figure \ref{F:arch-fig}:
\begin{enumerate}
\item User retrieves a list of applications hosted within the AHE from
the App Server Registry.

\item User invokes the Prepare operation in step 2 on the App Server
Factory.

\item As a result of the invocation of the Prepare operation, the App
Server Factory creates a new App WS-Resource representing the instance
of the application. The App Server Factory returns a WS-Addressing
\cite{wsaddressing} EndPoint Reference (EPR) to the client which the
client uses to communicate with the App WS-Resource.

\item The AHE client automatically transfers the user's input files to
the File Staging Area.

\item The user uploads his/her proxy credential to the MyProxy
server. This is then valid for one week by default.

\item The user starts the application instance and the AHE returns the
initial status of the application instance.

\item Once the App WS-Resource is created as a result of the Prepare
operation, its reference handle and associated properties are stored
in the App Instance Registry (see 3 above).

\item The AHE server creates one or more JSDL documents based on the
user's specification of the application instance and submits them to
GridSAM which then starts the application instance on the target grid
resource.

\item The App WS-Resource uses the proxy certificate for
authentication with the various grid resources that the user wishes to
use.

\item The input files are staged from the FSA to the target grid
resource on which the application instance will be executed.

\item The user can use the AHE client to invoke, monitor and terminate
commands on the App WS-Resource in order to check the status and, if
need be, to terminate the application instance.

\item Once the job finishes, the output files are transferred from the
grid resource to the FSA.

\item The user can use the AHE client to manually or automatically
download the output files from the FSA.

\item The user can query the App Instance Registry to review the
history of application instances.
\end{enumerate}

\end{document}